\documentclass[12 pt,reqno]{iopart}
\usepackage{amssymb}
\usepackage{epsfig}
\usepackage{color}
\newcommand{\be}{\begin{equation}}
\newcommand{\ee}{\end{equation}}
\newcommand{\ba}{\begin{eqnarray}}
\newcommand{\ea}{\end{eqnarray}}

\begin{document}

\title[Quasi-exact solutions for two interacting electrons]
{Quasi-exact solutions for two interacting electrons in
two-dimensional anisotropic dots}
\author{Przemys\l aw Ko\'scik and Anna Okopi\'nska\\
Institute of Physics, \'Swi\c{e}tokrzyska Academy\\
ul. \'Swi\c{e}tokrzyska 15, 25-406 Kielce, Poland}


\begin{abstract}

\noindent We present an analysis of the two-dimensional
Schr\"{o}dinger equation for two electrons interacting via Coulombic
force and confined in an anisotropic harmonic potential. The
separable case of $\omega_{y}:\omega_{x}=2$ is studied particularly
carefully. The closed-form expressions for bound-state energies and
the corresponding eigenfunctions are found at some particular values
of $\omega_{x}$. For highly-accurate determination of energy levels
at other values of $\omega_{x}$, we apply an efficient scheme based
on the Fr\"{o}benius expansion.
\end{abstract}
\maketitle

\section{Introduction}
The Hookean system composed of $N$ interacting particles trapped in
an external harmonic potential \be V(\textbf{r})= {m_{*}\over
2}[\omega_{x}^2 {x} ^2+\omega_{y}^2 {y} ^2+\omega_{z}^2{z}
^2]\label{trap}\ee has been considered first as a model of nucleus
in the early days of nuclear physics~\cite{nucl}. In the case the
two particles interact via central forces, the Hookean system enjoys
a particularly nice feature that the center-of-mass motion may be
separated out. Because of its relatively easy tractability, the
system has nowadays become a standard tool for testing quality of
many-body approximation methods in studying correlations between
electrons. Recently, the Hookean system attracted renowned interest,
since the progress in semiconductor technology has allowed the
formation of quantum dots (QDs), which are the systems of few
electrons confined on the scale of hundreds of nanometers in an
approximately harmonic potential~\cite{fab}. For the simplest QD,
consisting of two electrons ($N=2$), the Hamiltonian can be written
as

\begin{eqnarray} H=\sum_{i=1}^2[{\textbf{p}_{i}^2\over 2 m_{*}}
+V(\textbf{r}_{i})]
 +{e^2\over
\varepsilon_{*}|\textbf{r}_{2}-\textbf{r}_{1}|},~~~~\label{ham}\end{eqnarray}
where $m_{*}$ is the effective electron mass and $\varepsilon_{*}$
the effective dielectric constant. Introducing center of mass and
relative coordinates \be \textbf{R}={1\over
2}(\textbf{r}_{1}+\textbf{r}_{2})=(X,Y,Z),~\textbf{r}=\textbf{r}_{2}-\textbf{r}_{1}=(x,y,z),\ee
the above Hamiltonian is separated into
$H=H^{\textbf{R}}+H^{\textbf{r}}$, where the center of mass
Hamiltonian \be H^{\textbf{R}}={\textbf{P}^2\over
4m_{*}}+m_{*}(\omega_{x}^2 X^2+\omega_{y}^2Y^2 + \omega_{z} ^2
Z^2)\label{cm}\ee is exactly solvable. The problem is thus reduced
to the Schr\"{o}dinger equation for the relative motion \be
H^{\textbf{r}}\psi=\varepsilon^{\textbf{r}}\psi\label{rS},\ee
described by the Hamiltonian \be
H^{\textbf{r}}=-\hbar^2\nabla^2/m_{*} +
{m_{*}\over{4}}[{\omega}_{x}^2 x^2+{\omega}_{y}^2 y^2+{\omega}_{z}^2
z^2]+{e^2\over \varepsilon_{*}r},\label{ham2}\ee where
$r=|\textbf{r}|$.
 By a transformation $\hat{x}={x/
a_{*}},\hat{y}={y/ a_{*}}$,
 the relative motion  equation
is  rescaled to the  form
 \be[ -{\partial^2\over
\partial \hat{x}^2}-{\partial^2\over
\partial \hat{y}^2}-{\partial^2\over
\partial \hat{z}^2}+\hat{\omega}_{x}^2 \hat{x}^2 +\hat{\omega}_{y}^2 \hat{y}^2+\hat{\omega}_{z}^2 \hat{z}^2 +{1\over
{\sqrt{\hat{x}^2+\hat{y}^2+\hat{z}^2}}}]\psi=\hat{\varepsilon}^{\textbf{r}}\psi\label{Sch}\ee
where 
$\varepsilon^{\textbf{r}}= R^{*}\hat{\varepsilon}^{\textbf{r}}$,
$\omega_{x,y,z}/2=\hat{\omega}_{x,y,z}( \hbar/m_{*} a_{*}^2 )$ with
the effective Bohr radius $a_{*}=\hbar^2\varepsilon_{*}/m_{*} e^2$
and the effective Rydberg constant $R^{*}=m_{*} e^4/\hbar^2
\varepsilon_{*}^2 $. In the following, we skip the hats over spatial
coordinates $x,y,z$, frequencies $\omega_{x,y,z}$ and relative
energy $\varepsilon^{\textbf{r}}$.

In the case of some particular relation between the frequencies
$\omega_{x}$, $\omega_{y}$ and $\omega_{z}$, the problem of the
relative motion becomes easier to solve, and at some specific values
of frequencies it even appears quasi-solvable, i.e. one of the
states of the spectrum can be analytically determined.
The most studied example, important for describing QDs in a constant
magnetic field, is the case the two of confinement frequencies are
equal, e.g. $\omega_{x}=\omega_{y}=\omega$. For such axially
symmetric system there are three cases where the problem of the
relative motion~(\ref{Sch}) is  integrable: 1) $\omega_{z}=\omega$,
being separable in spherical coordinates, 2) $\omega_{z}:\omega=2$,
being separable in parabolic
 coordinates, and 3) $\omega_{z}:\omega=1/2$, which is one of
few examples of integrable problems that is generally not separable
\cite{Sim2}. The quasi-solvability of the case 1) has been known for
longtime~\cite{herschbach,taut}, and recently it has been
demonstrated~\cite{anali} that also in two other cases, one of the
bound-state solutions (not necessarily the ground-state) can be
obtained in a closed-form at some particular values of $\omega$.

Another example, which is important in studying two-dimensional (2D)
QDs, is that where one of the frequencies is much larger than the
other two, e.g. $\omega_{z}\rightarrow \infty$. The 2D approximation
is usually justified in practice, since QDs are realized by lateral
confinement of electron gas at the boundaries of semiconductor
nanostructures. The above problem is integrable in two cases: that
of $\omega_{x}=\omega_{y}$, and that of $\omega_{y}:\omega_{x}=2$
(which is equivalent to the case of $\omega_{y}:\omega_{x}=1/2$,
under the exchange of $\omega_{x}$ and $\omega_{y}$). In the first
case, the harmonic potential has circular symmetry and the problem
of the relative motion becomes separable in polar coordinates. This
case was much studied in the literature \cite{Merkt,Zhu1,wkb} and
the closed-form solutions for particular frequencies have been
derived \cite{tautB}. The influence of the anisotropy aspect ratio
$\omega_{x}/\omega_{y}$ on the QDs properties has been recently
studied in Ref. \cite{anis,anis1}, where the separability of the
case of $\omega_{y}:\omega_{x}=2$ in parabolic coordinates has been
pointed out. Here we shall study this integrable case in detail. The
main result of our paper is the demonstration that for particular
values of $\omega_{x}$ the exact solutions can be found in a closed
form. Although the property of quasi-solvability appears only at
particular confinement frequencies, it is very appealing, as the
knowledge of exact bound-state energies and the corresponding
eigenfunctions in a closed form can be used for extracting
information about the model and for testing the performance of
approximation methods. Therefore, we discuss the quasi-solvable
cases in detail, providing the values of confinement frequencies
$\omega_{x}$ as well as the corresponding energy eigenvalues and the
closed-form solutions for wave-functions. For determining the energy
spectrum at arbitrary values of $\omega_{x}$, we develop a numerical
scheme based on the Fr\"{o}benius method (FM), which was
successfully used for calculating bound state energies of
one-dimensional \cite{bar,alh,alh1} and spherically symmetric
potentials \cite{ko,zakrzewski}. The comparison of the bound-state
solutions determined within the FM with the exact ones allows us to
demonstrate how good is the performance of the numerical
approximation scheme.

The outline of our work is as follows. In section~\ref{2}, we give a
brief discussion of the spectrum in 2D case of motion (\ref{Sch2}),
paying a particular attention to the integrable case of
$\omega_{y}:\omega_{x}=2$. In section~\ref{3}, the quasi-solvable
cases are derived. In section~\ref{4}, the calculation within the FM
are described and a comparison with exact results is given. The
paper ends with concluding remarks in section~\ref{5}.

\section{ Theoretical study}\label{2}

In the case of laterally confined QD, the relative motion
equation~(\ref{Sch}) takes the form
 \be[ -{\partial^2\over
\partial {x}^2}-{\partial^2\over
\partial {y}^2}+{\omega}_{x}^2 {x}^2 +{\omega}_{y}^2 {y}^2+{1\over
{\rho}}]\psi(\mathbf{r})={\varepsilon}^{\textbf{r}}\psi(\mathbf{r}),\label{Sch2}\ee
where $\rho=\sqrt{x^2+y^2}.$ The singlet (s) and triplet (t) spin
states of the two-electron system correspond to symmetric and
antisymmetric spatial wave-functions, respectively. As the
centre-of-mass coordinate remains the same upon the interchange of
electrons, the symmetry requirement reduces to the symmetry of the
relative wave-function under inversion $\mathbf{r}\rightarrow
-\mathbf{r}$ . Because of the invariance of equation (\ref{Sch2}) to
reflections about the $x$- and $y$-axis, the $(x,y)$-parity of
spatial wave-functions is well-defined. The parity $(+,+)$ or($-,-)$
corresponds to spin singlet eigenfunctions, and the parity $(+,-)$
or $(-,+)$ to the spin triplet ones.

The simplest case of circular symmetry has been much studied in the
literature \cite{Merkt,Zhu1,wkb}. Here we discuss the other
integrable case, $\omega_{y}:\omega_{x}=2$. Introducing
two-dimensional parabolic coordinates through
 \be x=\eta_{1}\eta_{2},y={1\over
2}(\eta_{1}^2-\eta_{2}^2), \eta_{2}\geq 0 \label{parabolic}\ee the
Schr\"{o}dinger equation is transformed into the form  \be
H^{\textbf{r}}(\eta_{1},\eta_{2})\psi(\eta_{1},\eta_{2})=\varepsilon^{\textbf{r}}\psi(\eta_{1},\eta_{2})\label{parab}\ee
with the Hamiltonian given by
\begin{eqnarray} H^{\textbf{r}}(\eta_{1},\eta_{2})={1\over
\eta_{1}^2+\eta_{2}^2}[-{\partial^2\over
\partial\eta_{1}^2}-{\partial^2\over
\partial\eta_{2}^2}+{1\over 4}\omega_{y}^2(\eta_{1}^6+\eta_{2}^6)\nonumber\\
-{1\over
4}\eta_{1}^2\eta_{2}^2(\eta_{1}^2+\eta_{2}^2)(\omega_{y}^2-4\omega_{x}^2)+2]\label{hampar}\end{eqnarray}

Now, it is easy to see that by setting $\omega_{y}=2 \omega_{x}$ and
$\psi(\eta_{1},\eta_{2})=g_{1}(\eta_{1})g_{2}(\eta_{2})$ into
(\ref{hampar}) one gets  two ordinary differential equations of the
identical form \be [-{d^2\over
d\eta_{j}^2}-\varepsilon^{\textbf{r}}\eta_{j}^2+\omega_{x}^2
\eta_{j}^6+\beta_{j}]g_{j}(\eta_{j})=0, j=1,2\label{par}\ee that are
coupled by the condition $\beta_{1}+\beta_{2}=2$ to be satisfied by
the separation constants $\beta_{1}$ and $\beta_{2}$. The
integrability is explained by the fact that in the case of
$\omega_{y}: \omega_{x}=2$ an operator commuting with the
Hamiltonian~(\ref{hampar}) exists in the form \cite{anis} \be
\Lambda=\{\hat{L}_{z},\hat{p}_{x}\}+2\omega_{x}^2 y
x^2-{y\over\rho},\label{oper}\ee where the bracket $\{\}$ denotes
anticommutator. The operator $\Lambda$ commutes with the
$x-$reflection but anticommutes with the $y-$reflection operator,
and its eigenvalue is given by $\delta=(\beta_{1}-\beta_{2})/2$.
Therefore, the eigenstates corresponding to $\delta\neq 0$, i.e.
$\beta_{1}\neq\beta_{2}$, are  doubly degenerate with respect to the
sign of $\delta$, and those with $\delta=0$, i.e.
$\beta_{1}=\beta_{2}$, are nondegenerate.

The bound-states of the QD correspond to the values of
$\varepsilon^{\textbf{r}}$ and $\beta_{j}$ such that the functions
$\psi(\eta_{1},\eta_{2})$ vanish, as $|\eta_{1}|,\eta_{2}\rightarrow
\infty$. The eigenfunctions of (\ref{par})
 which vanish at infinity can be
ordered by the quantum number $n_{j}$ that counts its nodal points
on $-\infty< \eta_{j}<\infty$, and will be denoted by
$g^{(n_{j})}(\eta_{j},\beta_{j})$. The reflection symmetry of of
(\ref{par}) divides $g^{(n_{j})}(\eta_{j},\beta_{j})$ into two
parity types, depending on $n_{j}$ being even or odd.
 As the inverse coordinate
transformation
$$\eta_{1}=\frac{x}{|x|}\sqrt{\rho+y} \mbox{~~~~and~~~~}
\eta_{2}=\sqrt{\rho-y},$$  shows discontinuity of in the $\eta_{1}$
coordinate at positive $y$-axis, the condition emerges that the
solution of~(\ref{parab}) be a product of functions of the same
parity in $\eta_{1}$ and $\eta_{2}$, since only such products have
the appropriate continuity properties in the whole space. In the
case of $\delta=0$, which corresponds to the functions
$g(\eta_{j},\beta_{j})$ having the same number of nodal points, the
singlet state (s) with $(x,y)-$parity $(+,+)$ is represented by the
product of functions being even in $\eta_{1}$ and $\eta_{2}$, while
the triplet state (t) with $(-,+)$ is represented by the product of
odd functions. For $\delta \neq 0$ the solutions of~(\ref{parab})
with well-defined $(x,y)$-parity are easily constructed as \be
\psi_{(\pm)}^{(n_{1},{n_{2})}}(\eta_{1},\eta_{2})=g^{(n_{1})}(\eta_{1},\beta_{1})g^{(n_{2})}(\eta_{2},\beta_{2})\pm
g^{(n_{1})}(\eta_{2},\beta_{1})g^{(n_{2})}(\eta_{1},\beta_{2}),\label{wfdp}\ee
where the sign $(+)$ corresponds to the singlet/triplet state with
$(x,y)-$parity $(+,+)$/$(-,+)$, and the sign $(-)$ to the
triplet/singlet state with $(+,-)$/$(-,-)$ in the case of $n_{1}$
and $n_{2}$ being even/odd.

\begin{figure}[h]
\begin{center}
\epsfbox{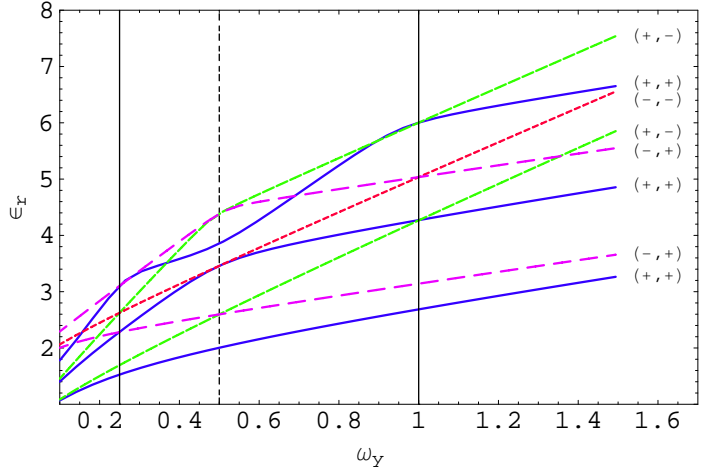}
\end{center}
 \caption{\label{fig1:beh}The
behaviour of the relative motion energy levels of a QD at
$\omega_{x}=0.5$ in function of $\omega_{y}$. The vertical lines
correspond to the cases $\omega_{x}:\omega_{y}=2$ and
$\omega_{y}:\omega_{x}=2$. The circularly symmetric case of
$\omega_{y}=\omega_{x}$ is marked by the broken line. }
\end{figure}

In order to visualize the manifestation of degeneracies discussed
above, we consider the behavior of bound-state energies for varying
anisotropy $\omega_{x}/\omega_{y}$. The numerical calculation have
been performed with the help of the linear Rayleigh-Ritz method
using the basis of two-dimensional harmonic oscillator eigenstates
("exact diagonalization"). The energies of low-lying states of the
relative motion, $\varepsilon^{\textbf{r}}$, are plotted in figure
\ref{fig1:beh}, for a fixed frequency $\omega_{x}=0.5$ in function
of the parameter $\omega_{y}$. It can be observed how the states
with spatial parity, $(+,-)$,$(-,+)$ become degenerate with those of
$(+,+),(-,-)$, respectively, in the case discussed above,
 i.e.
$\omega_{y}:\omega_{x}=2$, and in the case equivalent to it after
exchange $x\rightleftarrows y$, i.e. $\omega_{x}:{\omega_{y}}=2$.
The visible degeneracy of the spectrum at
$\omega_{y}=\omega_{x}=0.5$ corresponds to a circular symmetry of
the harmonic potential.

In the following we will discuss the spectrum of the anisotropic QD
in the case of $\omega_{y}:\omega_{x}=2$, which can be determined by
solving a coupled pair of one-dimensional equations (\ref{par}).  A
simple scheme based on the FM will be used for the numerical
determination of the solutions in the above case. The exact solution
will be provided for particular values of $\omega_{x}$.

\section{Exact solutions}\label{3}
Let us to come to main purpose of this paper which is the
demonstration that the problem of a 2D QD, confined by the potential
${\omega}_{x}^2 {x}^2 +{\omega}_{y}^2 {y}^2$, is quasi-solvable for
${\omega}_{y}:{\omega}_{x}=2$, i.e. that some of the solutions of
the relative motion equation (\ref{Sch2}) can be found algebraically
for certain values of $\omega_{x}$. We are looking for a solution of
the form
$\psi(\eta_{1},\eta_{2})=g(\eta_{1},\beta_{1})g(\eta_{2},\beta_{2})$,
where $\beta_{1}+\beta_{2}=2$, and $g(\eta_{j},\beta_{j})$ fulfils
the ordinary differential equation (\ref{par}). As the form of
Eq.\ref{par} is similar to that of the sextic anharmonic oscillator,
the quasi exact solutions can be represented as~\cite{quasi} \be
g(\eta_{j},\beta_{j})= e^{-{ {\omega_{x} \eta_{j}^4\over 4}}
}\sum_{i=0}^{\infty}a_{i}^{(j)}\eta_{j}^{2i+\nu},\label{ser}\ee
where $a_{0}^{(j)}\neq 0$, and $\nu = 0,1$ correspond to the even-
and odd-parity case, respectively. Substituting (\ref{ser}) into
(\ref{par}) and comparing the coefficients of like powers of
$\eta_{j}$,  one gets the three-term recursion relation for the
expansion coefficients $a_{i}^{(j)}$ in the form \be
A_{i}a_{i+1}^{(j)}+B_{i}^{(j)}a_{i}^{(j)}+C_{i}a_{i-1}^{(j)}=0, \label{rec}\ee where \ba
A_{i}&=&(2i+\nu+1)(2i+\nu+2)
\nonumber\\B_{i}^{(j)}&=&-\beta_{j}\nonumber\\
C_{i}&=&\varepsilon^{\textbf{r}}+(1-4i-2\nu)\omega_{x}\label{rec1}\ea
 with $a_{i}^{(j)}=0$ for $i<0$.
 It turns out that the QD equation (\ref{Sch2}) is exactly solvable
 if the series in the representations (\ref{ser}) for both
$g(\eta_{1},\beta_{1})$ and $g(\eta_{2},\beta_{2})$, terminate after
a finite number of terms. This gives the three conditions
\be\varepsilon^{\textbf{r}}=\omega_{x}(3+4n+2\nu)\label{en}\ee
\begin{eqnarray}
a_{n+1}^{(1)}(\omega_{x},\varepsilon^{\textbf{r}},\beta_{1})=0
\mbox{~~~~and~~~~}
a_{n+1}^{(2)}(\omega_{x},\varepsilon^{\textbf{r}},\beta_{2})=0,\label{con}\end{eqnarray}
where $n$ denotes the number of the highest non-vanishing term which
appears to be necessarily the same in both series. Using the
recurrence relation (\ref{rec}), the coefficients of the series may
be conveniently represented in the form
\begin{eqnarray}
a_{n+1}^{(j)}(\omega_{x},\varepsilon^{\textbf{r}},\beta_{j})={(-1)^{n+1}a_{0}^{(j)}\over
(2(n+1)+\nu)!
}P_{n}^{(j)}(\varepsilon^{\textbf{r}},\beta_{j},\omega_{x})=\nonumber\\{(-1)^{n+1}a_{0}^{(j)}\over
(2(n+1)+\nu)! }Det \pmatrix{ B_{0}^{(j)}&A_{0}&...&...&0\cr
C_{1}&B_{1}^{(j)}&A_{1}&...&0\cr 0&C_{2}&B_{2}^{(j)}&A_{2}&0\cr
...&...&...&...&...\cr0&...&C_{n-1}&B_{n-1}^{(j)}&A_{n-1}\cr
0&...&...&C_{n}&B_{n}^{(j)}\cr},j=1,2.\label{poly}\end{eqnarray} By
calculating the above determinant for a chosen value of $n$, the
polynomials $P_{n}^{(j)}$ may be easily determined and expressed in
terms of $\beta_{j}$ and $\varepsilon^{\textbf{r}}$ upon taking
(\ref{en}) into account. As an illustration we list here the first
four polynomials \be
P_{1}^{(j)}(\varepsilon^{{\textbf{r}}},\beta_{j})=
-{4\varepsilon^{{\textbf{r}}}(1+\nu)(2+\nu)\over
7+2\nu}+\beta_{j}^{2}\ee \be
P_{2}^{(j)}(\varepsilon^{{\textbf{r}}},\beta_{j})={4\varepsilon^{{\textbf{r}}}(16+\nu(13+3\nu))\beta_{j}\over
11+2\nu}-\beta_{j}^{3}\ee
\begin{eqnarray}
P_{3}^{(j)}(\varepsilon^{{\textbf{r}}},\beta_{j})={48{\varepsilon^{{\textbf{r}}}}^2(1+\nu)(2+\nu)(5+\nu)(6+\nu)\over
(15+2\nu)^2}-\nonumber\\{8{\varepsilon^{{\textbf{r}}}}(30+\nu(17+3\nu))\beta_{j}^{2}\over
15+2\nu}+\beta_{j}^4\end{eqnarray}
\begin{eqnarray}
P_{4}^{(j)}(\varepsilon^{{\textbf{r}}},\beta_{j})=-{16{\varepsilon^{{\textbf{r}}}}^2(2944+\nu(3404+\nu(1451+5\nu(50+3\nu))))
\beta_{j}\over
(19+2\nu)^2}+\nonumber\\+{40{\varepsilon^{{\textbf{r}}}}(16+\nu(7+\nu))\beta_{j}^{3}\over
19+2\nu}-\beta_{j}^5\end{eqnarray} The values of
$\varepsilon^{\textbf{r}},\beta_{1},\beta_{2}$, for which a
closed-form solution of a given parity $\nu$ exist, are determined
by solving the algebraic equations
$P_{n}^{(1)}(\varepsilon^{{\textbf{r}}},\beta_{1})=0$ and
$P_{n}^{(2)}(\varepsilon^{{\textbf{r}}},\beta_{2})=0$,
simultaneously with the imposition of $\beta_{1}+\beta_{2}=2$ at a
chosen value of $n$. The solution  may be explicitly determined by
calculating the coefficients $a_{i}^{(j)}$ in the functions
$g(\eta_{j},\beta_{j})$ via~(\ref{poly}) for $j=1,2$. We may notice
that for a fixed confinement potential with a specific value of the
frequency $\omega_{x}$ determined by (\ref{en}), there is only one
solution that is exactly known. This is a consequence of the fact
that the energy $\varepsilon^{{\textbf{r}}}$ enters equation
(\ref{en}) as a coefficient of the quadratic term in the potential,
which is in difference with the case of the sextic oscillator
Schr\"{o}dinger equation, where several solutions may be
analytically determined at specific values of the anharmonicity
parameter~\cite{quasi}.

\begin{figure}[h]
\begin{center}
\includegraphics[width=0.9\textwidth]{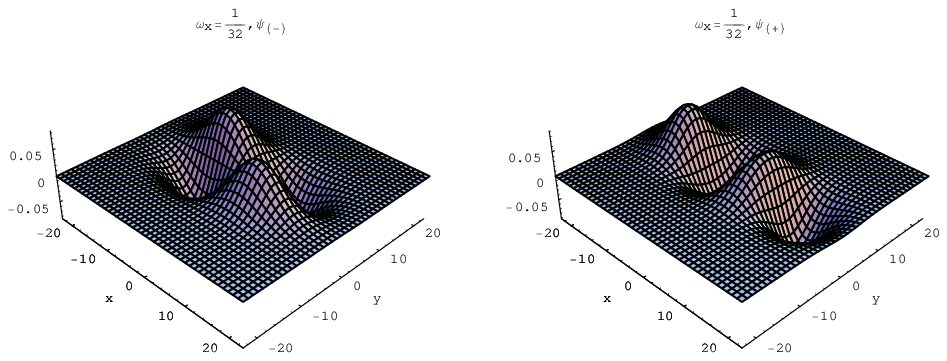}
\end{center}
 \caption{\label{fig2:beh}
The exact wave functions $\psi_{(\pm)}$ corresponding to the
degenerate level of the relative motion energy
$\varepsilon^{\textbf{r}}={13\over 32}$ }.
\end{figure}

\begin{figure}[h]
\begin{center}
\includegraphics[width=0.9\textwidth]{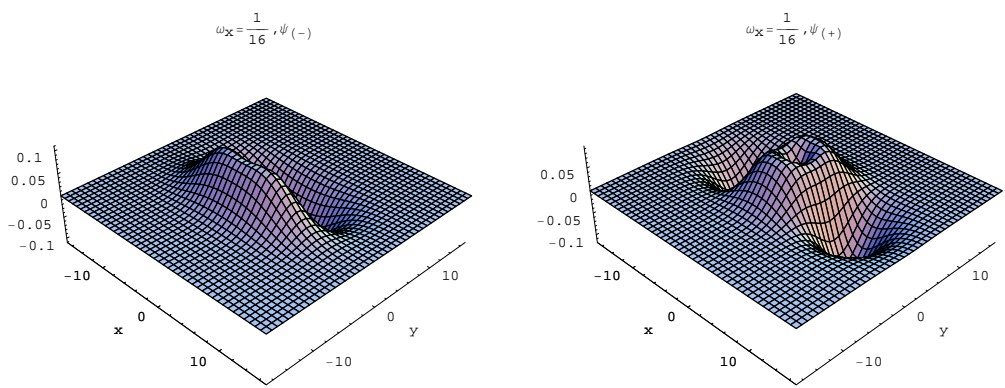}
\end{center}
 \caption{\label{fig3:beh}The exact wave functions $\psi_{(\pm)}$ corresponding to the degenerate level
 of the
relative motion energy $\varepsilon^{\textbf{r}}={11\over 16}$ }
\end{figure}

We consider here the lowest exact solutions of for the 2D asymmetric
QD generated through setting $n=1,2,3,4,...,etc$. In the simplest
case of $\nu=0$ and $n=1$, we have
 the equations
$P_{1}^{(j)}(\varepsilon^{\textbf{r}},\beta_{j})=-{8\varepsilon^{\textbf{r}}\over
7}+\beta_{j}^2=0$, which upon condition $\beta_{1}+\beta_{2}=2$, are
solved by  $\beta_{j}=1$, $\varepsilon^{\textbf{r}}={7/8}$,
corresponding to the functions $g^{(0)}(\eta_{j},1)=e^{-{
{{\eta_{j}^4}\over 32} }}(1+{\eta_{j}^2\over 2})$ that are nodeless.
This yields the normalized exact solution of (\ref{Sch2}) in
cartesian coordinates in the form \be \psi^{(0,0)}(x,y)={1\over
2\sqrt{(12\pi\sqrt{2}+\Gamma[-{1\over 4}]^2+2\Gamma[{1\over 4}]^2)}}
e^{-{x^2\over 16}-{y^2\over 8}}(1+{1\over 4}x^2+\rho),\ee which
corresponds to the ground-state. Further examples obtained in this
way, for both the ground and excited states, are given in Table
\ref{tab:table1} with the corresponding values of $\omega_{x}$,
$|\delta|$, and $\varepsilon^{\textbf{r}}$. In Figures
\ref{fig2:beh} and \ref{fig3:beh}, we have plotted the normalized
wave functions $\psi_{\pm}$, that correspond to the pairs of
degenerate states, obtained for $\omega_{x}=1/32$ and
$\omega_{x}=1/16$, respectively.

\begin{table}[h]
\begin{center}
\caption{\label{tab:table1} Confinement frequencies, $\omega_{x}$,
and the corresponding exact values of the relative motion energy,
$\varepsilon^{\textbf{r}}$, in the case of $\omega_{y}:\omega_{x}=2$
at a given parity $\nu$ for several values of $n$.} \lineup
\begin{tabular}{@{}ccccccccc}
\br $n$&$\nu$& ${{\omega_{x}}}$ &$spin$& \0 symmetry
parity&$|\delta|$ &
\0$\varepsilon^{\textbf{r}}$&$(n_{1},n_{2})$\\
\mr
 1&0&${1\over 8}$  &\0(s)&\0$(+,+)$&$0$ &${7\over 8}$&$(0,0)$\\
2&&${1\over 64}$  &\0(s)&\0$(+,+)$& $0$&${11\over 64}$&(0,0)\\
  &&${1\over 16}$  &\0(s),(t)&\0$(+,+),(+,-)$&$ 1$ &${11\over 16}$&(2,0)\\

 3&&${1\over 120}(5+ 2\sqrt{5})$  &\0(s)&\0$(+,+)$&$0$ &${1\over 8}(5+ 2\sqrt{5})$&$(2,2)$\\

 &&${1\over 120}(5- 2\sqrt{5})$  &\0(s)&\0$(+,+)$&$0$ &${1\over 8}(5- 2\sqrt{5})$&$(0,0)$\\

 &&${1\over 240}(5+\sqrt{5})$  &\0(s),(t)&\0$(+,+),(+,-)$&${1\over 2}(\sqrt{5}+ 1)$ &${1\over 16}(5+\sqrt{5})$&$(0,4)$\\

&&${1\over 240}(5-\sqrt{5})$  &\0(s),(t)&\0$(+,+),(+,-)$&${1\over 2}(\sqrt{5}- 1)$ &${1\over 16}(5-\sqrt{5})$&$(2,0)$\\
1&1&${1\over 24}$  &\0(t)&\0$(-,+)$&$0$ &${3\over 8}$&$(1,1)$\\
2&&${1\over 128}$  &\0(t)&\0$(-,+)$& $0$& ${13\over 128}$&$(1,1)$\\
&&${1\over 32}$  &\0(s),(t)&\0$(-,-),(-,+)$&$ 1$ &${13\over 32}$&$(3,1)$\\
\br
\end{tabular}
\end{center}
\end{table}

\eject

\section{Fr\"{o}benius method}\label{4}

Now we develop a numerical scheme based on the Fr\"{o}benius method
(FM) for obtaining the approximate energies of an asymmetric
QD~(\ref{Sch2}) in the integrable case of $\omega_{y}:\omega_{x}=2$.
The method consists of expanding the solution of a differential
equation into power series \cite {Fuchs}, and was originally applied
by Barakat and Rosner~\cite{bar} to compute the spectrum of
one-dimensional quartic oscillator confined by impenetrable walls at
$x= \pm R$. The energy eigenvalues of the system have been obtained
numerically as zeros of a function, calculated from its power series
representation. Moreover, it has been shown that the bound-state
energies of the confined system approach rapidly those of the
unconfined oscillator for increasing $R$ \cite{bar,alh,alh1}. Here
we show that solving the two separated equations (\ref{par}) with
the help of the power series representation (\ref{ser})
simultaneously with the condition $\beta_{1}+\beta_{2}=2$ enables us
to determine effectively the energy levels of the QD to very high
precision. With the boundary conditions imposed on the functions
$g_{j}(\eta_{j})$ at the points $\eta_{j}=R $, namely
$g(R,\beta_{j})=0$, the above problem reduces to finding common
zeros $[\delta,\varepsilon^{\textbf{r}}]$ of the functions
$g(R,1+\delta)$ and $g(R,1-\delta)$. The functions $g(R,1\pm\delta)$
can be calculated numerically with an arbitrary accuracy from the
power series representations  \be g(\eta_{j},\beta_{j})=
\sum_{i=0}^{K}b_{i}^{(j)}\eta_{j}^{2i+\nu},\label{serFM}\ee
truncated at a suitably high order $K$.
The obtained approximations to $g(\eta_{j},\beta_{j})$ are
polynomials in the variables $\varepsilon^{\textbf{r}}$ and
$\delta$, which can be taken to advantage in determining the
numerical values of their common zeros. Since the exact bound-states
of the QD correspond to the functions $g(\eta_{j},\beta_{j})$ that
vanish at $\eta_{j}\rightarrow \infty$, we expect the solution of
the parabolically confined problem to approach the desired solution
of the unconfined system, as the confinement range $R$ increases.
For the demonstration of convergence, we apply the above scheme to
the QD with $\omega_{x}=1/16$, for which the exact solution is known
(see Table \ref{tab:table1}). Table \ref{tab:table2} shows how the
numerical values $[\delta,\varepsilon^{\textbf{r}}_{FM}]$, obtained
for the confined system, approach the exact ones, as the truncation
order $K$ in the power series (\ref{serFM}) and the confinement
range $R$ increase.

\begin{table}[h]
\begin{center}
\caption{\label{tab:table2} Convergence of the energy eigenvalue and
the corresponding separation constant determined by the FM. The
underlined numbers agree with the exact results.} \lineup
\begin{tabular}{@{}cccccccc}
\br $K$& $R$&\0$\varepsilon^{\textbf{r}}_{FM}$&$\0\delta$\\
\mr
60 & 5 &\0\underline{0.687}4432520&\0$$0.999811134& \\
 70 &  &\0\underline{0.68750}24844&\0$$\underline{1.00000}7959& \\
90 &5.5  &\0\underline{0.6875000}837&\0$$\underline{1.000000}296& \\
100 &  &\0\underline{0.687500000}6&\0$$\underline{1.00000000}2& \\
120 &  &\0\underline{0.687500000}8&\0$$\underline{1.00000000}3& \\
140 & 6 &\0\underline{0.6875000000}&\0$$\underline{1.000000000}& \\

\br
\end{tabular}
\end{center}
\end{table}

\begin{figure}[h]
\begin{center}
\epsfbox{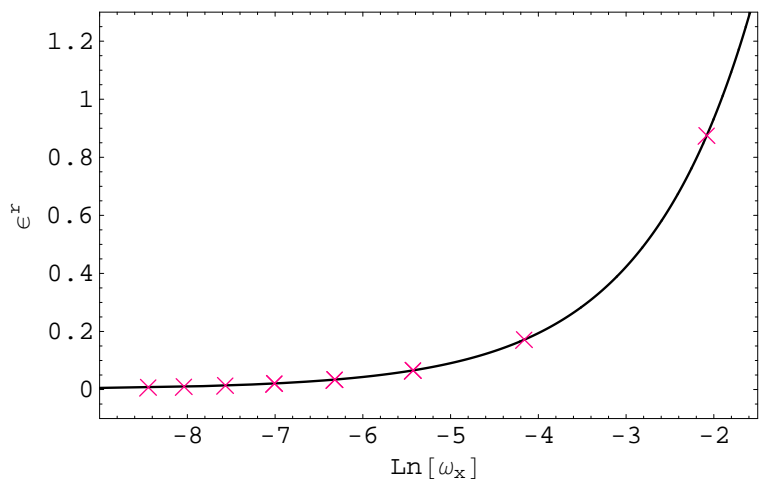}
\end{center}
 \caption{\label{fig4:beh}The relative motion ground
state energy $\varepsilon^{\textbf{r}}$ of a QD as a function of
Ln$\omega_{x}$($\omega_{y}:\omega_{x}=2$) determined by the FM. A
few values of energy, for which closed-form solutions exist are
marked with crosses.}
\end{figure}

We have also applied the FM for determining the precise values of
energies for the states that are not amenable to exact solution at
various $\omega_{y}:\omega_{x}=2$. The stability of numerical
results was achieved by increasing $K$ and $R$ until the energy
$\varepsilon^{\textbf{r}}_{FM}$ and the corresponding separation
eigenvalue $\delta$ stay fixed to the quoted accuracy, which
determines the energy of the relative motion of a QD to that
accuracy. The bound-state energies determined by the FM are listen
in Table \ref{tab:table3} up to 8-digit accuracy at particular
values of $\omega_{x}$. The states for which the exact solution
exists are marked bold, their energies can be shown to agree with
the exact values to practically unlimited accuracy. The plot of
ground state energy in function of Ln$\omega_{x}$ is displayed in
Fig. \ref{fig4:beh}, where we have marked a few of exact values of
energy, which correspond to the closed-form solutions discussed in
the previous Section.

We may notice that the other integrable case, that of circular
symmetry, where the problem reduces to the radial equation for the
relative motion, can be also treated within the FM. In that case the
solutions has a generalized power series representation and the
highly accurate values of energies can be determined with a small
computational effort by the scheme developed recently in our
previous work \cite{ko}.

\begin{table}[]
\caption{\label{tab:table3} Low-lying bound-state energies at
various confinement frequencies $\omega_{x}$, as determined by the
FM. The existing exact solution are marked bold.}
\begin{center}
\lineup
\begin{tabular}{cccccccc}
\br
$\omega_{x}$ &$\0\varepsilon^{\textbf{r}(+,+)}_{FM} $& $\0\varepsilon ^{\textbf{r}(-,+)}_{FM}$& $\0\varepsilon ^{\textbf{r}(+,+)(+,-)}_{FM}$ &$\0\varepsilon^{\textbf{r}(-,-),(-,+)}_{FM}$  \\
\mr \hline
1/64 &\textbf{0.17187500}&\00.172678376  &\00.220586977    &\00.225979324 \\
1/32 &0.293674143&  \00.297716776 & \00.386949414&\0\textbf{0.40625000}\\
1/24 &0.367590598&\0\textbf{0.37500000} &\00.490314824 &\00.521041536 \\
1/16 &0.505362736& \00.521827040 &\0 \textbf{0.68750000} &\00.743921450 \\
1/8 &\textbf{0.87500000}&\00.931629324 &\01.241576361 &\01.386478626 \\
1/6& 1.100931183&\01.191668135&\01.595014888 &\01.803819570\\
1/2&2.681851499&\03.142358616&\04.266819742  &\0 5.035425098\\
1 &4.773351606&\05.921649105&\08.099126926  &\09.762556072 \\
2 & 8.623556558&11.317113654& 15.569391839  &19.083841664 \\
\mr
\end{tabular}
\end{center}
\end{table}

On the example of a QD with confinement frequency $\omega_{x}={
1\over 64}$, we use the exact ground-state energy and the very
precise numerical values for higher bound-states determined by the
FM (see Table \ref{tab:table3}) to check the convergence of the
popular numerical methods applied directly to the two dimensional
Schr\"{o}dinger
equation
(\ref{Sch2}). First, we consider the Rayleigh-Ritz (RR) method with
the basis set of the two dimensional harmonic oscillator
eigenstates, calculating the approximate values of energies by
separate diagonalization of the matrix Hamiltonian corresponding to
the eigenfunctions with definite $(x,y)-$parity in a manner similar
to that used in Ref. \cite{anis}. In Table \ref{tab:table4} the
obtained values of energies $\varepsilon^{\textbf{r}}_{RR}$ are
presented for increasing dimension $D$ of the RR matrix. The same QD
example is used for testing the DVR (discrete variable
representation) method with the trigonometric basis  for the square
grid on a square whose sides have length $R$; $x_{i}=y_{i}=
-R+2Ri/N, i=1,...,N-1$ (for details see \cite{DVR1}). The numerical
values of energies $\varepsilon^{\textbf{r}}_{DVR}$ obtained by
diagonalization of the Hamiltonian matrix in point grid
representations with dimension $d=(N-1)^2$, are presented in Table
\ref{tab:table5} for several values of $R$ and $d$. We may see that
in both considered methods, the rough values of energies can be
determined with the use of small basis sets  but the convergence is
rather slow although the computational effort increases
quadratically with the size of the basis.

\begin{table}[]
\caption{\label{tab:table4} Convergence of low-lying states energy
eigenvalues calculated  by the Rayleigh-Ritz method for a QD with
$\omega_{x}=1/64$. The underlined numbers agree with the exact
result or that determined by the FM.}
\begin{center}
\lineup
\begin{tabular}{cccccccc}
\br
$D$  &$\0\varepsilon^{\textbf{r}(+,+)}_{RR} $& $\0\varepsilon ^{\textbf{r}(-,+)}_{RR}$& $\0\varepsilon ^{\textbf{r}(+,+)}_{RR}$ &$\0\varepsilon^{\textbf{r}(+,-)}_{RR}$ &$\0\varepsilon^{\textbf{r}(-,-)}_{RR}$&$\0\varepsilon^{\textbf{r}(-,+)}_{RR}$ \\
\mr

$ 25$ &\0\underline{0.1718}911&\0$$\underline{0.1726}898&
\0$$\underline{0.220}6744&\0$$\underline{0.220}6004&
\0$$\underline{0.2259}834&\0$$\underline{0.22}60368&\\

$ 64$  $ $&\0\underline{0.1718}856&\0$$\underline{0.1726}830&
\0$$\underline{0.220}6472&\0$$\underline{0.2205}919
&\0$$\underline{0.2259}803&\0$$\underline{0.2259}999&\\

$ 144$  $ $&\0\underline{0.1718}823&\0$$\underline{0.1726}804&
\0$$\underline{0.220}6274&\0$$\underline{0.22058}91&\0$$\underline{0.225979}6&\0$$\underline{0.2259}878&\\

$ 256$  $ $&\0\underline{0.1718}805&\0$$\underline{0.17267}95&
\0$$\underline{0.220}6169&\0$$\underline{0.22058}81&\0$$\underline{0.225979}5&\0$$\underline{0.2259}839&\\

 \mr
\end{tabular}
\end{center}
\end{table}

\begin{table}[]
\begin{center}
\caption{\label{tab:table5} Same as in Table \ref{tab:table4} but
calculated by the DVR method.} \lineup
\begin{tabular}{@{}cccccccc}
\br $d$&\0$R $&$ $&$\0$&$\0\varepsilon^{\textbf{r}}_{DVR}$&$\0$\\
\mr

$ 256$ & 30&\0$$\underline{0.171}9120& \0$$\underline{0.172}7763&\0$$\underline{0.220}6480&\0$$\underline{0.22}14837&\0$$\underline{0.22}60438&\0$$\underline{0.22}77602\\

$ 400$ &  &\0$$\underline{0.171}9294& \0$$\underline{0.172}7757&\0$$\underline{0.220}6258&\0$$\underline{0.22}15710&\0$$\underline{0.22}60337&\0$$\underline{0.22}77544\\

$ $ & 35&\0$$\underline{0.1718}335& \0$$\underline{0.1726}833&\0$$\underline{0.220}4264&\0$$\underline{0.220}6146&\0$$\underline{0.225}9909&\0$$\underline{0.22}60835\\

$576 $ & &\0$$\underline{0.1718}460& \0$$\underline{0.1726}824&\0$$\underline{0.220}4845&\0$$\underline{0.2205}998&\0$$\underline{0.2259}843&\0$$\underline{0.22}60752\\

$900$ & &\0$$\underline{0.1718}557& \0$$\underline{0.1726}815&\0$$\underline{0.2205}324&\0$$\underline{0.2205}924&\0$$\underline{0.2259}816&\0$$\underline{0.22}60703\\

$ $ &40&\0$$\underline{0.1718}483& \0$$\underline{0.17267}99&\0$$\underline{0.220}4538&\0$$\underline{0.2205}956&\0$$\underline{0.2259}819&\0$$\underline{0.2259}880\\

\br
\end{tabular}
\end{center}
\end{table}

\section{Concluding remarks}\label{5}
We discussed the case of a QD consisting of two interacting
electrons confined in an anisotropic harmonic potential $V(x,y)=
{m_{*}}(\omega_{x}^2 {x} ^2+ \omega_{y}^2 {y}^2)/2$. In the case of
$\omega_{y}: \omega_{x}=2$ we developed a numerical scheme based on
the Fr\"{o}benius method that allows an efficient determination of
the spectrum. Moreover, we have shown that a closed-form analytical
solutions for the relative motion can be obtained at particular
values of $\omega_{x}$. We illustrated using both the exact and the
very precise values of energies determined by FM for testing the
convergence of the Rayleigh-Ritz and DVR approximation methods. We
have shown that both the methods determine well the approximate
values of low-lying states energies but their convergence rates are
quite slow.

The exact values of energies and the corresponding wave functions
obtained in our work can be useful for investigating the performance
of other approximation methods in the case of the asymmetric
confinement. They can be also used to build the exact density
functional and testing the possible ways of constructing
approximations to it.

\end{document}